\documentclass[conference]{IEEEtran}
\IEEEoverridecommandlockouts
\usepackage{cite}
\usepackage{amsmath,amssymb,amsfonts}
\usepackage{algorithmic}
\usepackage{makecell}
\usepackage{graphicx}
\usepackage{subcaption}
\usepackage{textcomp}
\usepackage{xcolor}
\usepackage{siunitx}
\usepackage{booktabs}
\usepackage{multirow}
\def\BibTeX{{\rm B\kern-.05em{\sc i\kern-.025em b}\kern-.08em
    T\kern-.1667em\lower.7ex\hbox{E}\kern-.125emX}}
\begin{document}

\title{Unmixing The Crowd: Learning Persistent Speaker Representations from Mixture-Derived Multi-Speaker Embeddings
}

\author{\IEEEauthorblockN{Sidharth}
 \IEEEauthorblockA{
\textit{University of Michigan}\\
Ann Arbor, MI, USA \\
\textit{sidcs@umich.edu}}
\and
\IEEEauthorblockN{Meysam Asgari}
\IEEEauthorblockA{
\textit{Oregon Health \& Science University}\\
Portland, OR, USA \\
\textit{asgari@ohsu.edu}}
\and
\IEEEauthorblockN{Hao-Wen Dong}
 \IEEEauthorblockA{
\textit{University of Michigan}\\
Ann Arbor, MI, USA \\
\textit{hwdong@umich.edu}}
\and
\IEEEauthorblockN{Dhruv Jain}
\IEEEauthorblockA{
\textit{University of Michigan}\\
Ann Arbor, MI, USA \\
\textit{profdj@umich.edu}}

}

\maketitle

\begin{abstract}
We study whether persistent conversational speaker structure can be extracted directly from local overlapping speech mixtures. We propose a teacher-student framework that learns mixture-derived multi-speaker embeddings using only short overlapping segments and permutation-invariant latent supervision. Despite never being explicitly trained for speaker tracking, diarization, or conversational memory, the learned embedding space supports long-form speaker re-identification when combined with a lightweight online memory mechanism during inference. We additionally observe that the learned representation retains meaningful speaker structure under unseen overlap cardinalities. We further show that embeddings extracted from separation-first pipelines exhibit degraded clustering structure compared to embeddings predicted directly from mixtures. Finally, the learned embeddings remain effective for the downstream target speaker extraction task across multiple architectures. These findings suggest that local mixture-derived representations support persistent conversational speaker re-identification when combined with lightweight inference-time memory consolidation.
\end{abstract}

\section{Introduction}
Real-world conversations are highly dynamic — speakers enter and 
leave, overlap, pause, and reappear. For conversational AI systems, 
maintaining a stable representation of who is speaking is therefore 
as important as separating or enhancing speech itself.

Most existing approaches address this through speaker diarization, 
target speech extraction (TSE), or speech separation. Diarization 
systems track speakers using clustering pipelines or end-to-end 
neural approaches such as x-vectors~\cite{xvec} and 
EEND~\cite{eend}, with recent works exploring memory-aware 
representations for online tracking~\cite{ansd-ma-mse}. TSE systems 
condition enhancement models on enrollment examples~\cite{voicefilter, 
spex+}, while PIT-based separation systems~\cite{pit1,convtasnet} 
reconstruct multiple streams directly from the mixture. Despite strong 
performance, these methods share a common limitation: none are designed 
to maintain compact, reusable speaker representations that remain 
stable as speakers leave and re-enter a conversation.

We ask whether persistent speaker identity representations can emerge directly from local speech mixtures. Rather than relying on enrollment utterances, 
our model predicts candidate speaker embeddings directly from the 
input mixture. Embeddings trained only on short overlapping segments 
achieve speaker re-identification accuracy of 0.81 in simulated long-form conversations, despite never being trained for tracking or 
conversational memory. A lightweight online memory framework 
consolidates these local embeddings into persistent identities using 
embedding similarity and temporal confirmation, enabling re-identification 
across speaker absences. Direct mixture-derived embeddings consistently 
outperform separation-first pipelines (e.g. ~\cite{convtasnet}) on speaker 
identity preservation, and remain effective for downstream 
speaker-selective extraction.

The main contributions are:
\begin{enumerate}
    \item Mixture-derived multi-speaker embeddings trained only on 
    local overlapping mixtures support long-form speaker 
    re-identification despite never being trained for tracking or 
    diarization.
    \item A lightweight online memory framework enables low-switch-rate 
    identity persistence across conversations with more speakers than 
    observed locally during training.
    \item The learned representation retains meaningful structure under 
    unseen overlap cardinalities.
    \item The embeddings remain effective for downstream 
    speaker-selective extraction across multiple architectures.
\end{enumerate}

\section{System / Method}

\subsection{Mixture-Derived Multi-Speaker Embeddings}
Given a speech mixture containing multiple simultaneously active speakers, our goal is to directly estimate a compact set of latent speaker representations from the mixture itself. Rather than conditioning on externally provided enrollment utterances, the proposed framework predicts a fixed number of candidate speaker embeddings directly from the input audio. These embeddings are intended to capture speaker-specific identity information while remaining robust to overlap, background interference, and changing conversational context.

The overall framework follows a teacher-student framework formulation. A pretrained single-speaker embedding model acts as a teacher and produces clean speaker representations from isolated speech signals. A multi-speaker student model then learns to predict corresponding speaker embeddings directly from overlapping mixtures. 

The mixture $x_{\text{mix}} = \sum_{i=1}^{n_{sp}} s_i(t) + n(t)$ contains
$n_{sp}$ active speakers corrupted by additive noise $n(t)$.

 The student model predicts a fixed-size set of embeddings

$$\epsilon = \{ \varepsilon_{spk_1(GT)}, \varepsilon_{spk_2 (GT)},...,\varepsilon_{spk_K(GT)}\},
\\
\epsilon \in \mathbb{R}^{K \times D}$$ where $K$ corresponds to the maximum number of locally overlapping speakers observed during training. In this work, the student model is trained using two-speaker mixtures and therefore predicts two candidate speaker embeddings for each input segment. 

Importantly, the model is never trained using long-form conversational objectives, speaker tracking supervision, diarization losses or memory-based assignment strategies. Instead, the system only learns local mixture-level speaker representations from short overlapping segments. One of the central questions explored in this work is whether persistent speaker identity can nevertheless emerge from these local representations. 

\subsection{Teacher-Student Representation Learning}
To obtain robust speaker-identity supervision, we first train a teacher speaker embedding network using noisy single-speaker speech recordings. The teacher model is based on WavLM backbone~\cite{wavlm}, followed by attentive statistics pooling (ASP)~\cite{asp} and projection layers to produce fixed-dimensional speaker embeddings. 

For training the student model, for each clean source signal $s_i (t)$ before creating the mixture, the teacher network generates a normalized speaker representation 
\begin{equation}
    \varepsilon_{spk_i (GT)} = f_{teacher} (s_i), \varepsilon_{spk_i (GT)} \in \mathbb{R}^{D}
\end{equation}
 These teacher embeddings serve as latent identity targets for the student model. The teacher is trained with an ArcFace speaker verification objective~\cite{arcface}

The student network receives only the mixed audio waveform as input and predicts multiple candidate speaker embeddings simultaneously $$\hat{\varepsilon} = \{ \varepsilon_{spk_1}, \varepsilon_{spk_2},.., \varepsilon_{spk_K} \},\hat{\varepsilon} \in \mathbb{R}^{K \times D}$$

The student architecture shares the same WavLM backbone but replaces the single speaker pooling head with a multi-speaker embedding prediction module. This allows the model to directly disentangle multiple latent speaker representations from a single mixture segment.
\subsection{Permutation-Invariant Supervision}
Since the ordering of speakers inside a mixture is arbitrary, a fixed output ordering cannot be assumed during training. To address this ambiguity, we train the student with permutation-invariant embedding alignment strategy~\cite{pit1} (PIT applied to latent representations instead of speech signal outputs): for each mixture, we minimize the best assignment between predicted embeddings $\hat{\varepsilon}$ and teacher embeddings $\varepsilon$ computed from clean sources, using cosine distance. This enforces consistency with teacher space while allowing unordered set prediction.

\subsection{Online Memory Consolidation}
To study long-form speaker persistence, we investigate whether local mixture-derived multi-speaker embeddings can be consolidated into persistent conversational identities over time.
\subsubsection{Naive Memory Assignment (Baseline)}
As a lower bound, 
every unmatched embedding immediately creates a new speaker identity 
without temporal confirmation.
\subsubsection{Delayed Confirmation Method (Proposed)}
We introduce a lightweight online speaker memory mechanism operating entirely during inference. The memory framework incrementally consolidates local mixture-derived embeddings into persistent speaker representations over time. Conceptually, the proposed formulation is related to online speaker tracking and memory-aware diarization approaches~\cite{eend, ansd-ma-mse}, but differs in that the embeddings themselves are learned only from local overlapping mixtures rather than explicit long-form tracking supervision. 

For each incoming segment  $t$, the embedding model predicts a set of candidate speaker embeddings $\hat{\epsilon_t} = \{\hat{\varepsilon}_t^{(1)}, \hat{\varepsilon}_t^{(2)},...,\hat{\varepsilon}_t^{(K)} \}$, where $K$ denotes the maximum number of locally overlapping speakers supported by the model. The conversational memory bank maintains a set of persistent speaker centroids $M_t = \{m_1, m_2,..,m_{N_t}\}$ where $N_t$ denotes the number of memory identities accumulated up to time $t$. Memory association is performed using cosine similarity in the latent embedding space:
\begin{equation}
s(\hat{e}_t^{(k)}, m_j) = \frac{\hat{e}_t^{(k)}  m_j}{\lVert \hat{e}_t^{(k)} \rVert_2 \lVert m_j \rVert_2}    
\end{equation}
Each embedding is assigned to the memory identity with highest similarity. When the similarity exceeds a memory threshold $\tau_m$, the corresponding centroid is updated using running embedding aggregation:
\begin{equation}
    m_j' = \frac{n_j m_j + \hat{e}_t^{(k)}}{n_j + 1}
\end{equation}
where $n_j$ denotes the number of observations associated with memory identity $m_j$.

To reduce memory fragmentation caused by transient overlap artifacts or unstable local predictions, embeddings that do not match any existing memory identity (i.e., $\max_j s(\hat{e}_t^{(k)},m_j) < \tau_m$) are first stored as temporary pending candidates rather than immediately creating new speaker identities. A pending candidate is promoted into memory after $H$
consistent observations exceeding the same similarity threshold $\tau_m$. Pending candidates that are not re-observed within $A$ future segments are discarded.
As speakers leave and later re-enter the conversation, the memory bank enables consistent speaker re-identification through latent embedding similarity alone.

We further evaluate a single-speaker-guided update strategy built upon the delayed confirmation framework. In this variant, confirmed memory centroids are updated only using isolated single-speaker regions, while overlap regions are still used for memory assignment and candidate promotion. These embeddings are produced by the same dual-speaker embedding model, only the memory update policy is modified. The motivation is that embeddings extracted from isolated speech are generally more stable and less susceptible to overlap-induced distortion.
\subsection{Architecture Details}
\begin{figure}[t]
    \centering
    \begin{minipage}[b]{1\columnwidth}
        \centering
        \includegraphics[width=0.6\columnwidth]{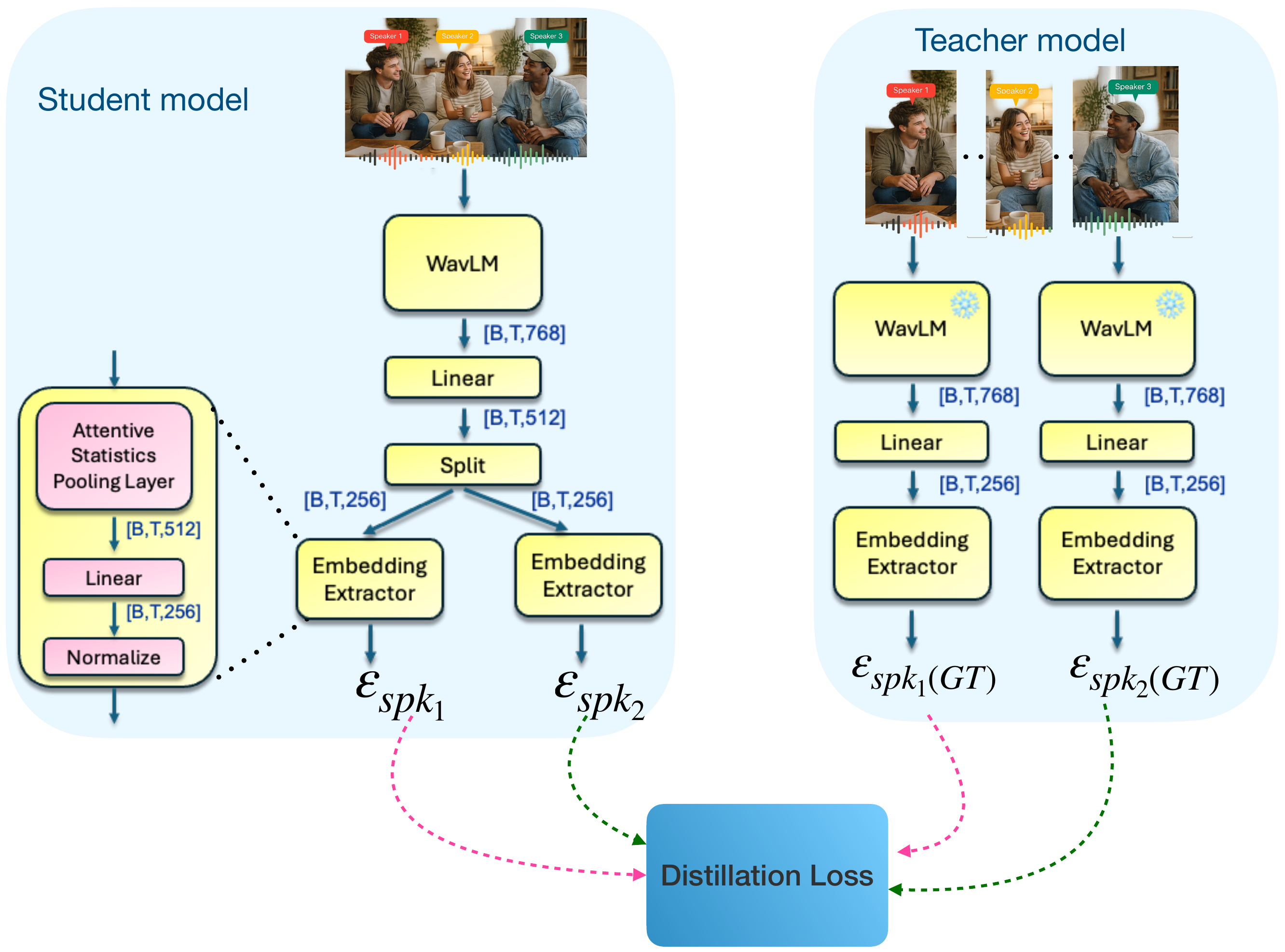}
        \subcaption{Teacher--student framework. The teacher defines a 
        single-speaker identity space; the student predicts an unordered 
        set of embeddings via permutation-invariant distillation.}
        \label{fig:architecture}
    \end{minipage}
    \hfill
    \begin{minipage}[b]{1\columnwidth}
        \centering
        \includegraphics[width=0.6\columnwidth]{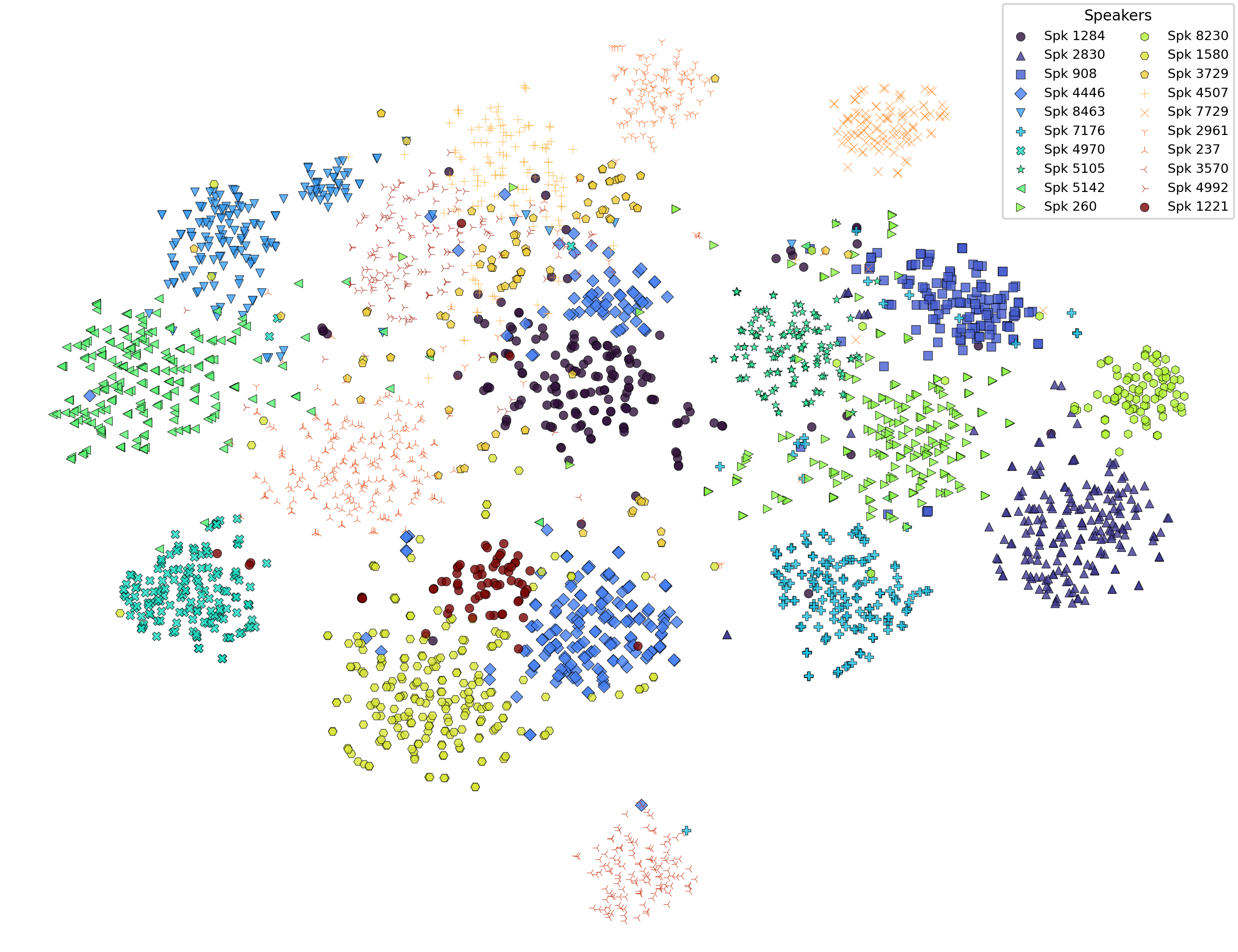}
        \subcaption{t-SNE visualization of embeddings from 20 speakers in the Libri2Mix test set. 
        Model: WavLM (FT) + 2EE w/ supervision.}
        \label{fig:TSNE}
    \end{minipage}
    \caption{System architecture and learned embedding structure.}
    \label{fig:arch_tsne}
\end{figure}

WavLM features are projected to D = 256 and pooled with ASP~\cite{asp, ecapa_tdnn} + $l_2$ normalization (Fig.~\ref{fig:architecture}). For the student model, we fine-tune only the top six layers of WavLM.

For delayed confirmation method, we use $\tau_m = 0.50$, $H = 2$, and $A = 8$.
\begin{table}[t]
    \centering
    \scriptsize
    \setlength{\tabcolsep}{2pt}
    \caption{Sensitivity of online memory consolidation to threshold $\tau_m$ 
    and reappearance window $H$, averaged over 150 simulated meetings.}
    \label{tab:sensitivity}
    \begin{tabular}{cccc}
        \toprule
        $\tau_m$ & $H$ &
        \textbf{Confirm Rate}$\uparrow$ &
        \textbf{Re-ID}$\uparrow$ \\
        \midrule
        \multirow{3}{*}{0.4}
            & 1 & 0.80 & 0.58 \\
            & 2 & 0.97 & 0.59 \\
            & 4 & \textbf{0.99} & \textbf{0.64} \\
        \midrule
        \multirow{3}{*}{0.5}
            & 1 & 0.35 & 0.62 \\
            & 2 & 0.61 & 0.59 \\
            & 4 & 0.77 & 0.55 \\
        \midrule
        \multirow{3}{*}{0.6}
            & 1 & 0.05 & 1.00$^{\dagger}$ \\
            & 2 & 0.12 & 0.94$^{\dagger}$ \\
            & 4 & 0.21 & 0.91$^{\dagger}$ \\
        \midrule
        \multicolumn{4}{p{0.8\columnwidth}}{$^{\dagger}$High Re-ID at $\tau_m{=}0.6$ 
        reflects severe under-confirmation ($<$21\% of speakers confirmed) rather 
        than genuine tracking quality. \textbf{Bold} denotes best joint tradeoff. 
        Paper uses $\tau_m{=}0.5$, $H{=}2$.} \\
        \bottomrule
    \end{tabular}
\end{table}
Table~\ref{tab:sensitivity} reports Re-ID accuracy and confirm rate across a range of $\tau_m$ and $H$ values. Performance degrades gracefully as $\tau_m$ increases, while $\tau_m$=0.4 with $H$=4 achieves the best joint tradeoff. We select $\tau_m$=0.5, $H$=2 as a conservative operating point that balances confirmation precision against speaker recall.


\section{Experimental Setup}
\label{sec: exp_setup}
\subsection{Training and Testing Data}
\label{sec:data}
We use the Libri2Mix and Libri3Mix pipelines~\cite{libri2mix} built from
LibriSpeech \texttt{train-clean-360}~\cite{librispeech}, modified to better approximate conversational overlap patterns. 
The proposed mixture-to-set encoder is not inherently tied to a fixed speaker count and could be extended with a dynamic-head when variable cardinality is required.
Specifically, we (i) enforce partial temporal overlap
by sampling an overlap ratio uniformly in 50--80\% and applying relative start-time
offsets between sources, and (ii) apply a speech-centric cropping step using
WebRTC-VAD to extract a 5\,s speech-containing segment from each source utterance prior
to mixing (with noise cropped/extended to match the resulting length). These mixtures
form the training and in-domain test data for all multi-speaker embedding models and
the TSE system. Background noise is added with SNR sampled from a 3-mode distribution
spanning $[-5,25]$\,dB (same scheme for train/test), using MUSAN (train) and WHAM!
(test). For the single-speaker teacher encoder, we instead corrupt individual
LibriSpeech \texttt{train-clean-360} utterances with noise from the Freesound portion
of MUSAN, again using the same SNR distribution as above, train on this noisy
single-speaker set, and evaluate on \texttt{test-clean} corrupted with WHAM! TT noise
using the same configuration.

\subsection{Downstream Speaker-Selective Extraction}
We evaluate our embeddings by conditioning three TSE back-ends: (i) pDCCRN~\cite{dccrn}
(with FiLM conditioning before the bottleneck LSTM), (ii) SpEx+~\cite{spex+},
and (iii) DPCCN~\cite{dpccn}. SpEx+ and DPCCN are taken from WeSep~\cite{wesep}
and adapted to accept our mixture-derived embeddings; for pDCCRN we use FiLM
modulation instead of embedding concatenation, following prior work on FiLM-based
conditioning~\cite{film, gfeller, itani_neural}.

Because the embedding model is trained with a permutation-invariant objective,
the predicted embeddings for a mixture have no fixed ordering. We therefore use
a teacher–student matching procedure only for supervision and evaluation:
each clean source is passed through a pretrained single-speaker embedding model
to obtain reference embeddings; for TSE training we randomly choose a target
source and select the predicted embedding that best matches its reference
(maximum cosine similarity). This provides an unambiguous training signal and,
at test time, enables offline bookkeeping to attribute scores to the
correct target.

\noindent\textbf{FiLM conditioning:}
We condition TSE models by predicting per-channel scale and bias from the selected embedding using a small MLP, and apply them to an intermediate encoder feature map (FiLM) before the bottleneck.

\subsection{Optimization and Evaluation}
\textbf{Loss Functions:} The teacher encoder is trained with ArcFace loss~\cite{arcface} 
($s{=}30.0$, $m{=}0.50$). The student is optimized with PIT 
cosine distance~\cite{pit1}. TSE back-ends follow their original 
loss formulations~\cite{spex+, dccrn, dpccn}.

\textbf{Metrics:} Following prior work on clusterable/self-supervised speaker embeddings~\cite{clustering_1, clustering_2, clustering_3, clustering_4}, we evaluate the embedding space using clustering accuracy, normalized mutual information (NMI), adjusted Rand index (ARI), Silhouette score, and a cosine-based separation measure  (mean cosine gap between same-speaker vs different-speaker pairs).

To evaluate long-form conversational consistency, we additionally report online speaker tracking metrics inspired by prior diarization and online speaker association literature~\cite{eend, ansd-ma-mse}. Specifically, we evaluate speaker re-identification accuracy, identity-switch rate, and conversational memory fragmentation. Re-identification accuracy measures how consistently recurring speakers are re-associated with the same memory identity across temporally separated conversational segments, while identity-switch rate quantifies the frequency of memory reassignment changes over time.

For TSE and separation, we report SI-SDR/SI-SDRi~\cite{sisdr}, PESQ~\cite{pesq}, STOI~\cite{stoi}, and DNSMOS P.835 (SIG/BAK/OVRL)~\cite{dnsmos,dnsmos2}.

\section{Results}
\subsection{Local Speaker Structure of Mixture-Derived Embeddings}
We first analyze the local structure of the learned embedding space before investigating long-form conversational persistence.

Table~\ref{tab:sp_clustering_metrics} summarizes the local clustering performance under noisy overlapping conditions.
\begin{table}[t]
    \centering
    \scriptsize
    \setlength{\tabcolsep}{2pt}
    \caption{Clustering performance for the 2-speaker embedding model on Libri2Mix test set with modifications described in Section~\ref{sec:data}. EE denotes the embedding extractor module (Fig.~\ref{fig:architecture}).}
    \label{tab:sp_clustering_metrics}

    \begin{tabular}{@{}lcccccc@{}}
        \toprule
        \textbf{Model} &
        \textbf{Params} &
        \textbf{Acc}$\uparrow$ &
        \textbf{Sep}$\uparrow$ &
        \textbf{NMI}$\uparrow$ &
        \textbf{Sil}$\uparrow$ &
        \textbf{ARI}$\uparrow$ \\
        \midrule

        \multicolumn{7}{c}{\textbf{Single Speaker Case}} \\
        \midrule

        ECAPA-TDNN~\cite{ecapa}
        & 15.9M & \textbf{97.00} & 0.45 & \textbf{0.98} & 0.22 & \textbf{0.94} \\

        WavLM (Frozen) + EE
        & 0.69M & 96.49 & 0.46 & \textbf{0.98} & 0.22 & 0.93 \\

        WavLM (FT) + EE
        & 47.9M & 93.59 & \textbf{0.51} & 0.96 & \textbf{0.27} & 0.89 \\

        \midrule
        \multicolumn{7}{c}{\textbf{Dual Speaker Case}} \\
        \midrule

        WavLM (Frozen) + K-means
        & -- & 7.35 & 0.008 & 0.05 & -0.07 & 0.005 \\

        WavLM (Frozen) + 2EE
        & 0.92M & 10.40 & 0.01 & 0.08 & -0.04 & 0.01 \\

        WavLM (FT) + 2EE
        & & & & & & \\
        \multicolumn{1}{l}{\quad -- w/ supervision}
        & 48.2M & \textbf{75.52} & \textbf{0.24} & \textbf{0.72} & 0.04 & \textbf{0.58} \\

        \multicolumn{1}{l}{\quad -- w/o supervision}
        & 48.2M & 38.33 & 0.07 & 0.27 & \textbf{0.15} & -0.03 \\

        ECAPA (Intermediate Features)
        & & & & & & \\
        \multicolumn{1}{l}{\quad + 2EE w/ supervision}
        & 44.9M & 73.67 & \textbf{0.24} & \textbf{0.72} & \textbf{0.06} & \textbf{0.58} \\

        \bottomrule
    \end{tabular}
\end{table}

In the single-speaker setting, the WavLM (Frozen) + EE teacher achieves performance comparable to the strong ECAPA-TDNN~\cite{ecapa} baseline, confirming that the frozen self-supervised backbone already provides a highly structured speaker representation space suitable for supervising multi-speaker embedding learning. Fine-tuning the upper WavLM layers slightly reduces clustering accuracy (93.6\%) but improves intra-/inter-speaker separation and Silhouette score, suggesting that adaptation to noisy speech conditions improves local embedding compactness while preserving overall speaker structure. Since these gains are relatively modest, we use the computationally cheaper WavLM (Frozen) + EE configuration as the teacher model in all subsequent experiments.

Frozen WavLM features fail to disentangle concurrent speakers under 
overlap, with clustering accuracy dropping to 7.35\% and near-zero 
NMI and ARI. Adding a dual-head extractor on a frozen backbone yields 
only marginal gains. Fine-tuning the upper transformer layers with the 
proposed PIT-based supervision restores speaker structure directly from 
noisy overlapping mixtures, confirming that multi-speaker 
disentanglement requires explicit mixture-aware supervision rather than 
emerging naturally from single-speaker representations. Replacing WavLM 
with ECAPA-based intermediate features produces comparable structure, 
suggesting the supervision framework is largely backbone-agnostic. 
Removing the teacher-guided objective degrades clustering across all 
metrics, highlighting the importance of permutation-invariant teacher 
alignment.

Overall, these results demonstrate that while modern self-supervised speech representations contain rich speaker information, they are insufficiently disentangled under overlap-heavy conditions. The proposed mixture-aware supervision framework enables the emergence of structured multi-speaker latent representations directly from noisy overlapping speech mixtures. Fig.~\ref{fig:TSNE} further visualizes the learned embedding space using t-SNE projections. To better understand whether this behavior is unique to direct mixture-derived representation learning, we additionally compare against separation-first pipelines where source separation is performed first followed by a downstream speaker embedding extraction in Section~\ref{separation_first}.


\subsection{Overlap Cardinality Generalization}

To investigate the sensitivity of the learned representation to overlap cardinality, we evaluate models trained under two-speaker and three-speaker overlap conditions on both matched and mismatched test sets. Table~\ref{tab:cardinality_generalization} summarizes the results.

Performance is highest under matched training and evaluation conditions. However, substantial speaker structure remains under cardinality mismatch. In particular, the two-speaker model retains meaningful clustering performance when evaluated on three-speaker mixtures, achieving an ARI of 0.44 and NMI of 0.60 despite never observing three-speaker overlap during training. Interestingly, the three-speaker model does not significantly outperform the two-speaker model on three-speaker mixtures, suggesting that representation quality rather than output cardinality appears to be the primary performance bottleneck.
\begin{table}[t]
    \centering
    \scriptsize
    \setlength{\tabcolsep}{4pt}
    \caption{Generalization across overlap cardinalities.}
    \label{tab:cardinality_generalization}

    \begin{tabular}{@{}lcccccc@{}}
        \toprule
        \textbf{Train} &
        \textbf{Test} &
        \textbf{Acc}$\uparrow$ &
        \textbf{Sep}$\uparrow$ &
        \textbf{NMI}$\uparrow$ &
        \textbf{Sil}$\uparrow$ &
        \textbf{ARI}$\uparrow$ \\
        \midrule

        2EE & 2-spk & \textbf{75.52} & \textbf{0.24} & \textbf{0.72} & \textbf{0.04} & \textbf{0.58} \\

        2EE & 3-spk & 64.30 & 0.21 & 0.60 & 0.03 & 0.44 \\

        3EE & 2-spk & 66.95 & 0.23 & 0.66 & \textbf{0.04} & 0.50 \\

        3EE & 3-spk & 60.77 & 0.20 & 0.58 & 0.03 & 0.41 \\

        \bottomrule
    \end{tabular}
\end{table}
\subsection{Comparison Against Separation-First Pipelines}
\label{separation_first}








\begin{table}[t]
    \centering
    \scriptsize
    \setlength{\tabcolsep}{2pt}
    \caption{Comparison against separation-first pipelines on Libri2Mix test set with modifications described in Section~\ref{sec:data}.}
    \label{tab:clustering_after_sep}

    \begin{tabular}{@{}lccccc@{}}
        \toprule
        \textbf{Method} &
        \textbf{Acc}$\uparrow$ &
        \textbf{Sep}$\uparrow$ &
        \textbf{NMI}$\uparrow$ &
        \textbf{ARI}$\uparrow$ &
        \textbf{Sil}$\uparrow$ \\
        \midrule

        ConvTasNet + ECAPA-TDNN
        & 11.00 & 0.014 & 0.076 & 0.011 & -0.02 \\

        WavLM (FT) + 2EE
        & \textbf{75.52} & \textbf{0.24} & \textbf{0.72} & \textbf{0.58} & \textbf{0.04} \\

        \bottomrule
    \end{tabular}
\end{table}

Beyond waveform-level separation quality, we ask whether a conventional
``separate-then-embed'' pipeline preserves a usable embedding space: Table~\ref{tab:clustering_after_sep} compares the embedding quality metrics between separate-then-embed pipeline and our proposed method. Embeddings extracted from ConvTasNet outputs exhibit poor clustering structure (Acc$\approx$11\%, NMI$\approx$0.08, ARI$\approx$0.011), suggesting that separation artifacts can distort the teacher embedding geometry. This is expected: ConvTasNet is optimized for waveform reconstruction under a separation objective, not for preserving speaker-identity cues in the teacher embedding space. To isolate this effect, we interpolate between clean references and separated outputs, $\tilde{s}(\alpha)=(1-\alpha)s_{\text{clean}}+\alpha s_{\text{sep}}$, after aligning $s_{\text{sep}}$ to
$s_{\text{clean}}$ via optimal gain and small-lag time-shift correction; clustering performance degrades monotonically
with $\alpha$ confirming that separation artifacts progressively perturb the
embedding space.

\subsection{Long-Form Conversational Memory}
To study long-form conversational persistence, we simulate meeting-like conversational mixtures containing approximately 10 unique speakers with dynamically varying activity patterns, speaker re-entry events, silence regions, and overlap-heavy interactions. At any given time, the maximum local overlap is restricted to two active speakers to remain consistent with the experiment settings.
\subsubsection{Naive Memory Assignment Leads to Severe Fragmentation}
Table~\ref{tab:memory_configs} summarizes the resulting conversational tracking behavior under direct online assignment.

Under this setting, the conversational memory fragments despite the conversation containing only ten unique speakers. In particular, the system produces an average of 33.20 inferred conversational memory identities, overestimating the true number of active speakers. This behavior is accompanied by high identity-switch rate of 0.70 and relatively poor re-identification consistency. 

Interestingly, the local embedding space itself remains moderately speaker-structured even under this unstable memory regime, as reflected by the ARI score. This suggests that the primary failure mode is not purely local embedding quality, but rather instability in long-form conversational identity consolidation. 

\begin{table}[t]
\centering
\scriptsize
\setlength{\tabcolsep}{2pt}
\caption{Conversational memory behavior under different memory policies and embedding representations.}
\label{tab:memory_configs}

\begin{tabular}{@{}lccccc@{}}
\toprule
\textbf{Configuration} &
\textbf{Confirm} &
\textbf{Est./True} &
\textbf{Re-ID}$\uparrow$ &
\textbf{Switch}$\downarrow$ &
\textbf{ARI}$\uparrow$ \\
&
\textbf{Rate} &
\textbf{Spk} &
& \textbf{Rate}
&
\\
\midrule

Direct Online Assignment
& 1.00 & 33.2/10 & 0.50 & 0.70 & 0.32 \\

Single-Speaker Guided Updates
& 0.45 & 7.90/10 & 0.75 & 0.37 & \textbf{0.43} \\

Conservative Memory Consolidation 
&  &  &  &  &  \\

\multicolumn{1}{l}{\quad -- Frozen WavLM + K-means}
& 0.89 & 3.40/10 & 0.67 & 0.52 & 0.004 \\

\multicolumn{1}{l}{\quad -- ConvTasNet + ECAPA}
& 0.98 & 2.00/10 & 0.66 & 0.47 & 0.01 \\

\multicolumn{1}{l}{\quad -- ECAPA + 2EE}
& 0.05 & 1.90/10 & 0.57 & \textbf{0.16} & \textbf{0.40} \\

\multicolumn{1}{l}{\quad -- WavLM (FT) + 2EE}
& 0.49 & 6.75/10 & \textbf{0.81} & 0.29 & 0.36 \\

\bottomrule
\end{tabular}
\end{table}

\subsubsection{Delayed Confirmation Stabilizes Conversational Identity}

The proposed delayed-confirmation framework stabilizes long-form conversational identity assignment compared to direct online memory creation. Under the conservative memory consolidation strategy, the number of inferred conversational speaker identities decreases dramatically from 33.20 to 6.75, while speaker re-identification accuracy improves from 0.50 to 0.81. Identity switch rate is simultaneously reduced from 0.70 to 0.29, indicating improved long-form speaker consistency.

In single-speaker-guided framework, since embeddings extracted from isolated speech are typically more stable and less corrupted by overlap interference, this strategy improves local clustering compactness and produces the highest ARI score among all evaluated configurations. However, reducing updates from overlap regions also slightly decreases conversational adaptability, leading to moderately increased identity switching compared to the more conservative memory consolidation strategy.

\subsubsection{Effect of Embedding Representation on Memory Consolidation}
To isolate the effect of the learned representation from the memory mechanism itself, we evaluate several alternative embedding strategies under the identical conservative memory consolidation policy. Table~\ref{tab:memory_configs} shows that memory behavior depends strongly on the underlying embedding representation. Frozen WavLM features clustered using K-means and embeddings extracted from separation-first pipelines both exhibit poor conversational structure, resulting in high identity-switch rates and weak clustering consistency. ECAPA-based dual-head embeddings produce strong local clustering quality but confirm only a small fraction of candidate speakers, leading to severe speaker under-discovery. In contrast, the proposed mixture-derived embeddings provide the best overall balance between memory coverage, speaker inventory recovery, and re-identification accuracy. These results suggest that the observed long-form conversational behavior cannot be attributed solely to the memory heuristic, but also depends on the structure of the learned embedding space.
\subsubsection{Speaker Lifetime and Re-Identification Analysis}
\begin{figure}[htbp]
\centerline{\includegraphics[width=\linewidth]{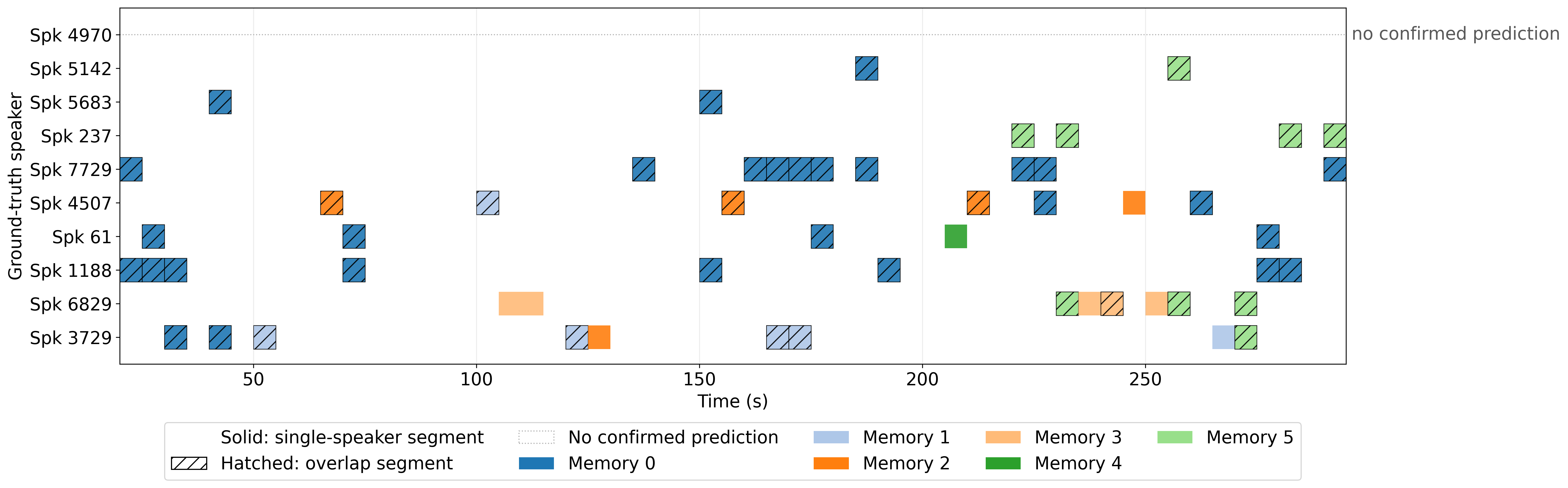}}
\caption{Conversational memory assignments across a representative 
long-form meeting. Each row corresponds to a ground-truth speaker. 
Colors indicate assigned memory identities, while hatched segments 
denote overlapping speech regions.}
\label{fig:lifetime_plot}
\end{figure}

Fig.~\ref{fig:lifetime_plot} shows memory assignments over time 
for a representative simulated conversation. Speakers are often reassigned to the same memory identity after extended absences, while identity inconsistencies cluster near overlap-heavy 
regions where local predictions are inherently less stable — 
consistent with the quantitative results in 
Table~\ref{tab:memory_configs}.

\subsection{Zero-Shot Transfer to Real Conversations}


        

\begin{table}[t]
    \centering
    \scriptsize
    \setlength{\tabcolsep}{1pt}
    \caption{Zero-shot tracking and diarization performance on the VoxConverse test set (English subset) without domain adaptation. All methods use the same online memory consolidation policy.}
    \label{tab:voxconverse_generalization}
    \begin{tabular}{@{}lcccccc@{}}
        \toprule
        \textbf{Method} &
        \textbf{Re-ID} $\uparrow$ &
        \textbf{Switch} $\downarrow$ &
        \textbf{NMI} $\uparrow$ &
        \textbf{ARI} $\uparrow$ &
        \textbf{Est./True} &
        \textbf{DER (\%)} $\downarrow$ \\
         &
          &
        \textbf{Rate}  &
         &
         & \textbf{Spk}
         & 
       \\
        \midrule
        Frozen WavLM + K-means &
        0.64 &
        0.44 &
        0.14 &
        0.05 &
        3.17 / 5 &
        56.50 \\
        ConvTasNet + ECAPA &
        0.83 &
        0.23 &
        0.10 &
        0.07 &
        1.71 / 5 &
        45.30 \\
        ECAPA(Intermediate Features) &
         &
         &
         &
         &
         &
         \\
        \multicolumn{1}{l}{\hspace{1em} + 2EE} &
        0.76 &
        \textbf{0.18} &
        \textbf{0.71} &
        \textbf{0.66} &
        4.98 / 5 &
        55.60 \\
        WavLM (FT) + 2EE &
        0.79 &
        0.27 &
        0.47 &
        0.36 &
        7.00 / 5&
        38.90 \\
        \midrule
        Pyannote~\cite{pyannote}$^{*}$ &- & -& -& -& -& \textbf{8.9}\\
        Softformer~\cite{softformer}$^{*}$ &- & -& -& -& -& 13.8\\
        Softformer-v2~\cite{softformer-v2}$^{*}$ &- & -& -& -& -& 15.1\\
        \bottomrule
    \end{tabular}
    \vspace{2pt}
    {\raggedright\scriptsize $^{*}$Results for pyannote 3.1, Sortformer, and Sortformer v2 reported 
from \cite{lanzendorfer2025benchmarking} on VoxConverse, systems 
trained with domain-specific supervision and not directly comparable 
to our zero-shot setting.}
\end{table}

To evaluate the generalization capabilities and robustness of our mixture-derived embeddings under real-world domain shifts, we conduct a zero-shot evaluation on the VoxConverse dataset~\cite{voxconverse}. Crucially, all methods are evaluated directly on this dataset without any fine-tuning or acoustic domain adaptation, relying entirely on representations learned from synthetic LibriSpeech mixtures.

As shown in Table~\ref{tab:voxconverse_generalization}, different embedding representations exhibit markedly different conversational tracking behaviors. Frozen WavLM features clustered using K-means produce weak speaker structure and poor tracking performance, while embeddings extracted from a separation-first pipeline achieve strong re-identification accuracy but under-estimate the number of active conversational speakers. ECAPA-based dual-head embeddings produce the strongest local clustering metrics (NMI: 0.71, ARI: 0.66), and recover approximately the correct number of speakers, indicating that the proposed training paradigm is not tied to a particular backbone architecture.

Despite being trained exclusively on synthetic LibriSpeech mixtures, the proposed representation achieves 38.9\% DER on VoxConverse without domain adaptation, compared to supervised systems trained on in-domain data achieving 8.9–15.1\%. The results demonstrate that mixture-derived embeddings learned from local overlap supervision can generalize beyond the training domain and remain useful for conversational speaker tracking without domain adaptation.
\subsection{Downstream Target Speech Extraction}
\begin{table}[t]
    \centering
    \scriptsize
    \setlength{\tabcolsep}{1pt} 
    \caption{Comparison of different TSE model architectures on Libri2Mix test set with modifications described in Section \ref{sec:data}.}
    \label{tab:model_compars_tse}
    \begin{tabular}{@{}lccccccc@{}}
        \toprule
        \textbf{Model}&  \textbf{Params}  & \textbf{STOI} $\uparrow$ & \textbf{PESQ} $\uparrow$ & \textbf{SIG} $\uparrow$ & \textbf{BAK} $\uparrow$ & \textbf{OVRL} $\uparrow$ & \textbf{SI-SDRi}$\uparrow$  \\
        \midrule

\multicolumn{8}{c}{\textbf{Dual Speaker Case}} \\
\midrule
Noisy & - & 0.74 & 1.18 & 2.98 & 2.69 & 2.33 & -\\
        pDCCRN~\cite{pdccrn}   &  &  &  &  &  &  &  \\
        \multicolumn{1}{l}{\quad -- no embedding} 
                & 3.7M & 0.75 & 1.19 & 3.18 & 2.32 & 2.18 & 1.0 \\
        \multicolumn{1}{l}{\quad -- concat} 
                & 3.7M & 0.78 & 1.37 & 2.62 & 3.21 & 2.16 & 5.8 \\
        \multicolumn{1}{l}{\quad -- FiLM} 
                & 3.8M & 0.82 &1.54 & 3.04 & 3.74 & \textbf{2.67} & 9.6 \\

        SpEx+~\cite{spex+} & 9.6M &  0.85 & 1.73 & 2.83 & 2.97 & 2.24 & 11.72\\
        DPCCN~\cite{dpccn} &  &  &  &  &  &  &  \\
         \multicolumn{1}{l}{\quad -- Student emb} & 6.4M &  0.87 & 1.82 & \textbf{3.06} & \textbf{4.00} & \textbf{2.67} & 12.83\\
        \multicolumn{1}{l}{\quad -- Teacher emb (FiLM)} 
                & 6.4M & \textbf{0.89} &\textbf{1.95} & 3.00 & 3.96 & 2.58 & \textbf{13.50 }\\
    
        \bottomrule
    \end{tabular}

\end{table}

Table~\ref{tab:model_compars_tse} compares several TSE architectures on noisy Libri2Mix mixtures using the proposed mixture-derived embeddings for conditioning.

DPCCN achieves the strongest overall performance, obtaining the best STOI, PESQ, and SI-SDRi scores while remaining relatively lightweight (6.4M parameters). 

We further observe that student-derived embeddings remain competitive with teacher-conditioned systems despite being inferred directly from overlapping mixtures without isolated enrollment utterances. This indicates that the proposed embedding space preserves sufficiently disentangled speaker structure for downstream controllable extraction tasks even under overlap-heavy noisy conditions.

Overall, these results suggest that the learned embeddings capture not only locally clusterable speaker information, but also robust speaker-specific conditioning cues that generalize effectively to downstream speech extraction architectures.

\section{Conclusion}
We studied whether persistent conversational speaker structure 
can emerge from local overlapping speech mixtures, without 
enrollment utterances, long-form tracking supervision, or 
conversational memory losses. While modern self-supervised 
representations fail to disentangle concurrent speakers under 
overlap, a permutation-invariant teacher-student supervision 
framework recovers this structure, yielding locally clusterable 
embeddings directly from noisy mixtures. Combined with a 
lightweight inference-time memory consolidation framework, 
these embeddings support speaker re-identification at 0.81 
accuracy with an identity-switch rate of 0.29 — despite never 
being optimized for tracking or diarization. Direct 
mixture-derived embeddings consistently outperform 
separation-first pipelines on speaker identity preservation, 
and remain effective for downstream target speaker extraction 
across multiple architectures. \textbf{Limitations:} The 
framework assumes fixed overlap cardinality and degrades as 
simultaneous speaker count increases; acoustically similar 
speakers are occasionally merged into the same memory identity.

\vspace{12pt}

\end{document}